\documentclass{iitpressproc}
\usepackage{graphicx}
\usepackage{amssymb}

\usepackage[sort&compress]{natbib}

\bibpunct{[}{]}{,}{n}{}{}

\begin{document}

\title{Large x physics:  recent results and future plans}

\author{Roy J. Holt
\address{Physics Division, Argonne National Laboratory, Argonne, IL, 60439}\\[2ex]
}

\maketitle

\begin{abstract}
The valence region is exceedingly important in hadron physics since this region not only defines a hadron but also is an excellent discriminator of nucleon structure models.  Present uncertainties in light quark  distribution functions at high $x$ could also impact high energy physics.  Here we will describe a new generation of experiments at Jefferson Lab that is aimed at the high $x$ region of the nucleon.  It is noted that the proposed Electron Ion Collider could explore the high $x$ regime.
\end{abstract}

\section{Introduction}

During the past four decades there has been an enormous effort to determine the parton distribution functions (PDFs) of the most stable hadrons: neutron, proton and pion \cite{Holt:2010vj}.  The behavior of the PDFs on the far valence region (Bjorken-$x> 0.5$) is of particular interest because this domain defines a hadron.
Recognizing the importance of the far valence region, a new generation of experiments, focused on $x\gtrsim 0.5$, is planned at the Thomas Jefferson National Accelerator Facility (JLab), and under examination in connection with Drell-Yan studies at the Fermi National Accelerator Facility (FNAL) \cite{Reimer:Figure} and a possible Electron Ion Collider (EIC), in China or the USA.
Theoretical calculations \cite{Roberts:2013mja,Chang:2013pq,Cloet:2013tta,Chang:2013nia} have moved far beyond the simple parameterization of PDFs.  These computations have emphasized the importance of nonperturbative QCD.  The importance of these calculations was illustrated by the pion's valence-quark PDF, $u_v^\pi(x)$, where a failure of QCD was suggested following a leading-order analysis of $\pi N$ Drell-Yan measurements \cite{Conway:1989fs} and with a lack of soft gluon resummation corrections \cite{Aicher:2010cb}.  On the other hand, a series of QCD-connected calculations  \cite{Hecht:2000xa,Wijesooriya:2005ir,Aicher:2010cb,Nguyen:2011jy} subsequently established that the leading-order analysis was misleading, so that $u_v^\pi(x)$ may now be seen as a success for the nonperturbative approach in QCD.  Finally, the high-$x$ region could impact high energy physics since low momentum transfer and high $x$ evolves to high momentum transfer and low $x$. This talk will focus on three main areas of high-$x$ research at JLab:  First, the planned measurements of the $F_2^n/F_2^p$ and $d/u$ ratios as well as present status;  Secondly,  measurements of the longitudinal spin asymmetries for the proton and neutron, while the third, the planned measurements of the transversity distributions of the nucleons.

\section{Experimental Status}

\suppressfloats[t]
\subsection{$F_2^n/F_2^p$ and $d/u$ ratios}

\begin{figure}
\hspace*{0.5in}\includegraphics[angle=-90,width=1.0\textwidth]{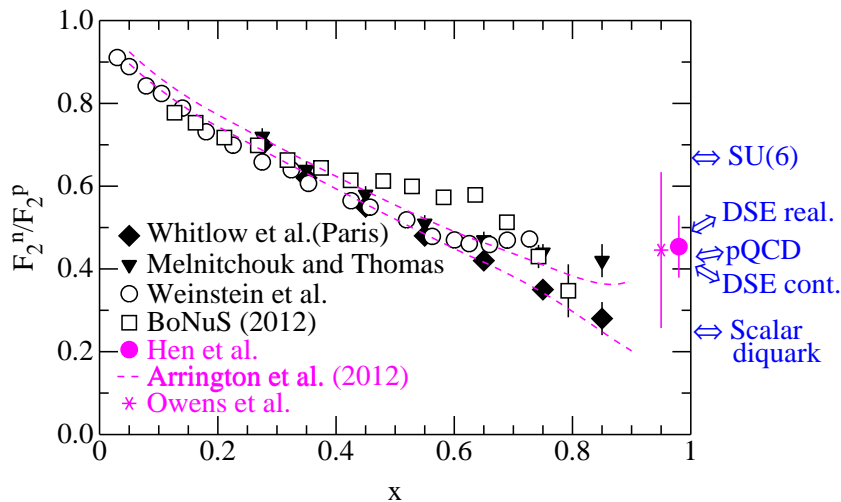}
\caption{\label{f2nratio} $F_2^n/F_2^p$ as a function of $x$.  Results from five extraction methods are shown \protect\cite{Whitlow:1991uw,Melnitchouk:1995fc,Arrington:2008zh,Weinstein:2010rt,Hen:2011rt,Arrington:2011qt,Owens:2012bv} along with selected predictions.  Adapted from ref.~\cite{Roberts:2013mja}.}

\end{figure}

There is considerable theoretical interest in the ratio of $F_2^n/F_2^p$ and $d/u$ at very high $x$.  The ratio as $x \rightarrow 1$ is sensitive to the model of the nucleon.  One can see from the right side of Fig.\,\ref{f2nratio} that the $F_2^n/F_2^p$ or $d/u$ ratio varies substantially among the models.
The ratio of the neutron structure function, $F_2^n$, to the proton structure function, $F_2^p$, is particularly interesting.  Within the parton model at very high x:
\begin{equation}
\label{F2nF2pratio}
\frac{F_2^n}{F_2^p} \stackrel{x \simeq 1}{=} \frac{1 + 4(d_v/u_v)}{4 + (d_v/u_v)}.
\end{equation}

Thus a measurement of the neutron and proton structure functions at large-$x$ provides a determination of the $d_v/u_v$ ratio.  However, while proton and deuteron DIS data are well measured at reasonably high $x$, the extraction of the neutron structure function at very high $x$ from DIS data on the deuteron is problematic.  The central difficulty is that the extraction of $F_2^n/F_2^p$ at high $x$ is sensitive to the poorly known high-momentum components of the deuteron wave function \cite{Arrington:2011xs}.

To see this, we note that many extractions of the neutron-proton structure function ratios have been performed \cite{Whitlow:1991uw,Melnitchouk:1995fc,Arrington:2008zh,Weinstein:2010rt,Hen:2011rt,Arrington:2011qt,Owens:2012bv}.  They are summarised in Fig.\,\ref{f2nratio}, with the three most recent inferences indicated by the points with error bars near $x=1$: there is a large uncertainty in the ratio for $x \gtrsim 0.6$.  (See also Fig.\,25 in Ref.\,\cite{Holt:2010vj}.)  New experimental methods are necessary in order to place tighter constraints on the far valence domain.  A primary goal is to empirically eliminate two of the three quite different theoretical predictions.

Two new experiments \cite{Bueltmann:2006,Petratos:2006} should provide data up to $x \approx 0.85$.  Since much of the uncertainty can be traced to the poorly known short-range part of the deuteron wave function, the JLab BoNuS Collaboration has performed \cite{Baillie:2011za} an experiment where a very low energy spectator proton from the deuteron can be detected in coincidence with a DIS event from the neutron in the deuteron.  In this way, one can restrict the data to a region where the well-known long-range part of the deuteron wave function dominates the process.  The central difficulty with this experiment is detecting the very low energy proton of about 150 MeV/c or less.  This requires extremely thin target and detector components. An interesting variant of this approach is to use an EIC with, e.g., an 8$\,$GeV electron beam impinging on a deuteron beam of 30$\,$GeV/nucleon.  The forward going energetic proton would be detected at very small angles in coincidence with a DIS event from the neutron.  Simulations suggest that this should be feasible \cite{Accardi:2010}.

Another method is to perform deep inelastic scattering from the mirror nuclei $^3$He and $^3$H over a broad range in $x$ \cite{Afnan:2000uh,Bissey:2000ed,Afnan:2003vh,Petratos:2006,Holt:2011zz}.  Theoretical calculations indicate that nuclear effects cancel to a high degree in extracting the $F_2^n/F_2^p$ ratio from these two nuclei.  This experiment would also be useful in determining the EMC effect in the mass-three system \cite{Hen:2013oha}.  Although providing a tritium target for JLab is straightforward\cite{Brajuskovic:2013ymh}, it is not trivial.

Finally, parity violating DIS can avoid the problem encountered with neutrons bound in nuclei.  Parity-violating DIS from the proton is sensitive to the $d/u$ ratio \cite{Souder:2010}.  Plans include measuring the $d/u$ ratio up to an $x$ of 0.7.


\subsection {Longitudinally polarized deep inelastic scattering}

\begin{figure}
\hspace*{0.3in}\includegraphics[angle=-90,width=.8\textwidth]{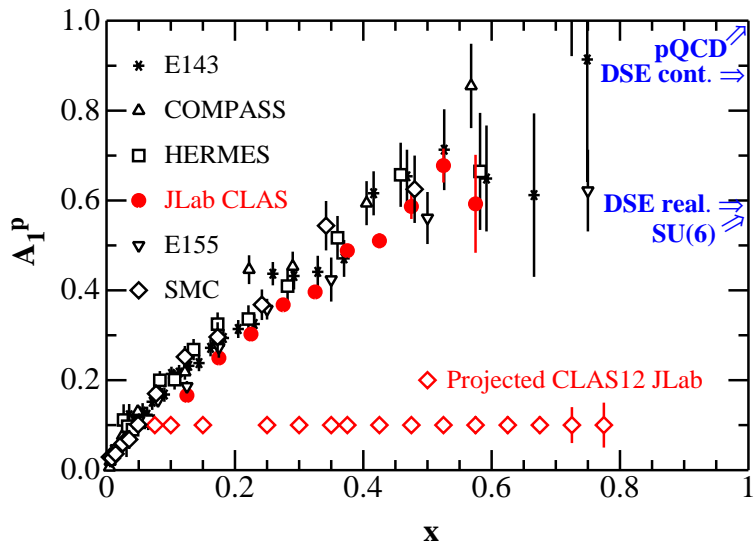}
\caption{\label{a1p}
Existing and projected measurements of the proton's longitudinal spin asymmetry as a function of $x$ (statistical errors only), along with selected predictions.  Adapted from ref.~\cite{Roberts:2013mja}.}
\end{figure}

It is evident from right side of Fig.\ref{a1p} that measurements of the longitudinal asymmetries in DIS provide an important constraint on models of nucleon structure.  Numerous experiments and extractions aimed at determining nucleon longitudinal spin structure functions have been performed\cite{Aidala:2012mv}.

Existing measurements of $A_1^p$ are summarised in Fig.\,\ref{a1p}.  Unfortunately, these data are not of sufficient accuracy and high enough $x$ to discriminate among the models.  As indicated in Fig.\,\ref{a1p}, however, a new JLab experiment \cite{kuhn:2006} will extend the results up to $x \approx 0.8$ with a projected error that promises a significant constraint on the models.

\begin{figure}
\hspace*{0.3in}\includegraphics[clip,angle=-90,width=.8\textwidth]{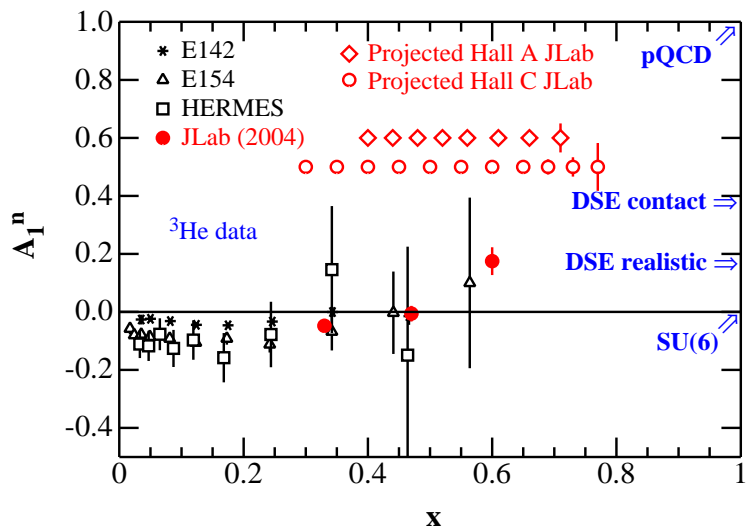}
\caption{\label{a1n} Existing and projected measurements of the neutron's longitudinal spin asymmetry as a function of $x$ (statistical errors only), along with selected predictions. N.B.\ We only display $A_1^n$ data obtained from polarised $^3$He targets.  Adapted from ref.~\cite{Roberts:2013mja}.}
\end{figure}

The status of existing data for $A_1^n$ is shown in Fig.\,\ref{a1n}.  The data extend only to $x \approx 0.6$ and place little constraint on descriptions of the nucleon.  New experiments proposed at JLab \cite{Zheng:2006,Liyanage:2006} are expected to provide results up to $x \approx 0.75$, as indicated in Fig.\,\ref{a1n}.  These new results should permit discrimination between the pQCD model and other predictions.

\subsection{transversity distributions}

The transversity distribution function is a chiral odd, T-even function that is accessible in Drell-Yan interactions and in semi-inclusive DIS (SIDIS).  The transversity is a measure of the quark transverse polarization in a transversely polarized nucleon.  A feature of the transversity distribution function is that the quark-antiquark and gluon seas do not contribute.  The nucleon tensor charge, an intrinsic property such as axial or vector charge, can be determined from the transversity distribution function, $h_1(x)$.  The tensor charges are important for calculation a permanent electric dipole moment of a nucleon.  In SIDIS, the asymmetry arises from the product $h_1(x)$ and the Collins fragmentation function, $H_1^\perp(x)$.  Fortunately, the Collins fragmentation function has been measured at Belle so that the $h_1(x)$ can be isolated.  A recent analysis \cite{Anselmino:2013vqa} of all transversity and Collins fragmentation function data has been performed and the tensor charges have been determined \cite{Prokudin:2013iza}.  At present the extracted value of the u-quark tensor charge is
$0.39^{+0.18}_{-0.12}$ and the d-quark tensor charge: $-0.25^{+0.30}_{-0.10}$.  The uncertainties in these values are generally too large to discriminate among the models.  However, a new generation of experiments would be expected to reduce these uncertainties by about an order of magnitude.
Transversity experiments for the proton \cite{Avakian:2011,Gao:2011} using a polarized hydrogen target and for the neutron \cite{Cates:2009,Gao:2010} {\it via} transversely polarized $^3$He targets have been planned for JLab at 12 GeV.  Data of such accuracy would add an important constraint on the available models.  An important recent finding is that within the framework of DSE, the inclusion of an axial diquark has a pronounced effect on the u-quark tensor charge.

\section{Summary}
An understanding of hadrons in terms of QCD is an essential goal of nuclear physics and would be a great contribution to science in general.  A vigorous program aimed at the high $x$ domain is planned at JLab at 12 GeV.  The goal of this program is to provide new data for the d/u ratio and the longitudinal spin asymmetries at very high $x$ as well as for the tensor charges of the nucleons.  These new results will provide unprecedented constraints on the models of the nucleon.

\section*{Acknowledgments}
I am grateful to C. D. Roberts, J. Arrington, C. Keppel, S. Kuhn and X. Zheng for fruitful discussions.
This work was funded by the
Department of Energy, Office of Nuclear Physics, contract no.~DE-AC02-06CH11357.



\bibliographystyle{apsrev}
\bibliography{Holt_ismd13_new}


\end{document}